**Exploring orbital-charge conversion mediated by interfaces with $CuO_x$ through spin-orbital pumping**

E. Santos[1], J. E. Abrão[1], A. S. Vieira[2], J. B. S. Mendes[2], R. L. Rodríguez-Suárez[3], and A. Azevedo[1]

*[1] Departamento de Física, Universidade Federal de Pernambuco, 50670-901, Recife, Pernambuco, Brazil.*
*[2]Departamento de Física, Universidade Federal de Viçosa, 36570-900, Viçosa, Minas Gerais, Brazil.*
*[3]Facultad de Física, Pontifícia Universidad Católica de Chile, Casilla 306, Santiago, Chile.*

We explore the impact of different materials on orbital charge conversion in heterostructures with a naturally oxidized copper capping layer. Introducing a thin layer of $CuO_x$ (3nm) to the yttrium iron garnet (YIG)/W heterostructure resulted in a notable decrease in signal when employing the spin pumping technique (SP). This contrasts with prior findings in YIG/Pt, where the addition of $CuO_x$ (3nm) led to a significant signal enhancement. Conversely, the introduction of the same $CuO_x$ (3nm) layer to YIG/Ti(4nm) structure showed no change in the SP signal. This lack of change is attributed to the fact that Ti, unlike Pt, does not generate an orbital current at the Ti/$CuO_x$ interface due to its weaker spin-orbit coupling. Notably, incorporating the $CuO_x$(3nm) layer into Si/Py(5nm)/Pt(4nm) structures resulted in a substantial increase in the spin pumping signal. However, in Si/$CuO_x$(3nm)/Pt(4nm)/Py(5nm) structures, the signal exhibited a decrease. Finally, we applied a phenomenological model of the spin (orbital) Hall effect in YIG/heavy metal systems to refine our data. These discoveries have the potential to advance research in the innovative field of orbitronics and contribute to the development of new technologies based on spin-orbital conversion.

## 1. Introduction

Exploring the properties of electrons beyond the spin degree of freedom has significantly expanded the scope of spintronics. Although the use of spin has traditionally dominated the field, orbital angular momentum (OAM) emerges as a crucial player in electron transport within solids. Theoretical predictions and recent experimental results [1-8] have shown that it is possible to have a flow of OAM perpendicular to a charge current, even with the quenching of OAM in solids or in materials with a weak SOC. This effect, known as the orbital Hall effect (OHE), has the property of being independent of the SOC, thus being considered a more fundamental effect [1], while the spin Hall effect (SHE) [9,10,11] assumes a secondary role. Similar to the spin-to-charge current conversion effects, orbital current can be converted to charge current within the bulk via OHE [12] or by interface phenomena [13]. The latter being known as orbital Rashba-Edelstein-like effect (OREE) [14-17].

The OREE was initially proposed as a theoretical concept and subsequently confirmed experimentally on surfaces with negligible SOC [15,18,19]. Rashba-like coupling between the vectors $\vec{L}$ and $\vec{k}$, leads to both orbital-dependent energy splitting and chiral OAM texture in k-space. Despite their similarities, the mechanisms governing the spin Rashba-Edelstein effect (SREE) and OREE differ significantly. The distinguishing feature of OREE is its independence from spin-orbit coupling, which makes it possible to manipulate orbital properties driven by interface effects. In Fig. 1(a), the upward charge current $\vec{J}_C$, generates a perpendicular spin current (represented by the red symbols) induced by SHE, and a perpendicular orbital current (represented by the oriented circles) induced by OHE. Because of the significant strength of the SOC, both the spin and orbital currents intertwine to form a perpendicular spin-orbital current $\vec{J}_{L,S}$. Fig. 1(b) illustrates the inverse effect, wherein an upward current $\vec{J}_{L,S}$ induces a current $\vec{J}_C$ through the inverse SHE (ISHE) and inverse OHE (IOHE). While Figs. 1(a) and 1(b) depict the occurrence of SHE and OHE within the volume, the interfacial counterparts are illustrated in Figs. 1(c) and 1(d). Fig. 1(c) illustrates the generation of a perpendicular orbital current $\vec{J}_L$, generated by the flow of an interfacial charge current $\vec{J}_C$. On the other hand, Fig. 1(d) illustrates the inverse effect of that illustrated in Fig. 1(c). A bulk orbital current $\vec{J}_L$ will generate a perpendicular interfacial charge current $\vec{J}_C$.

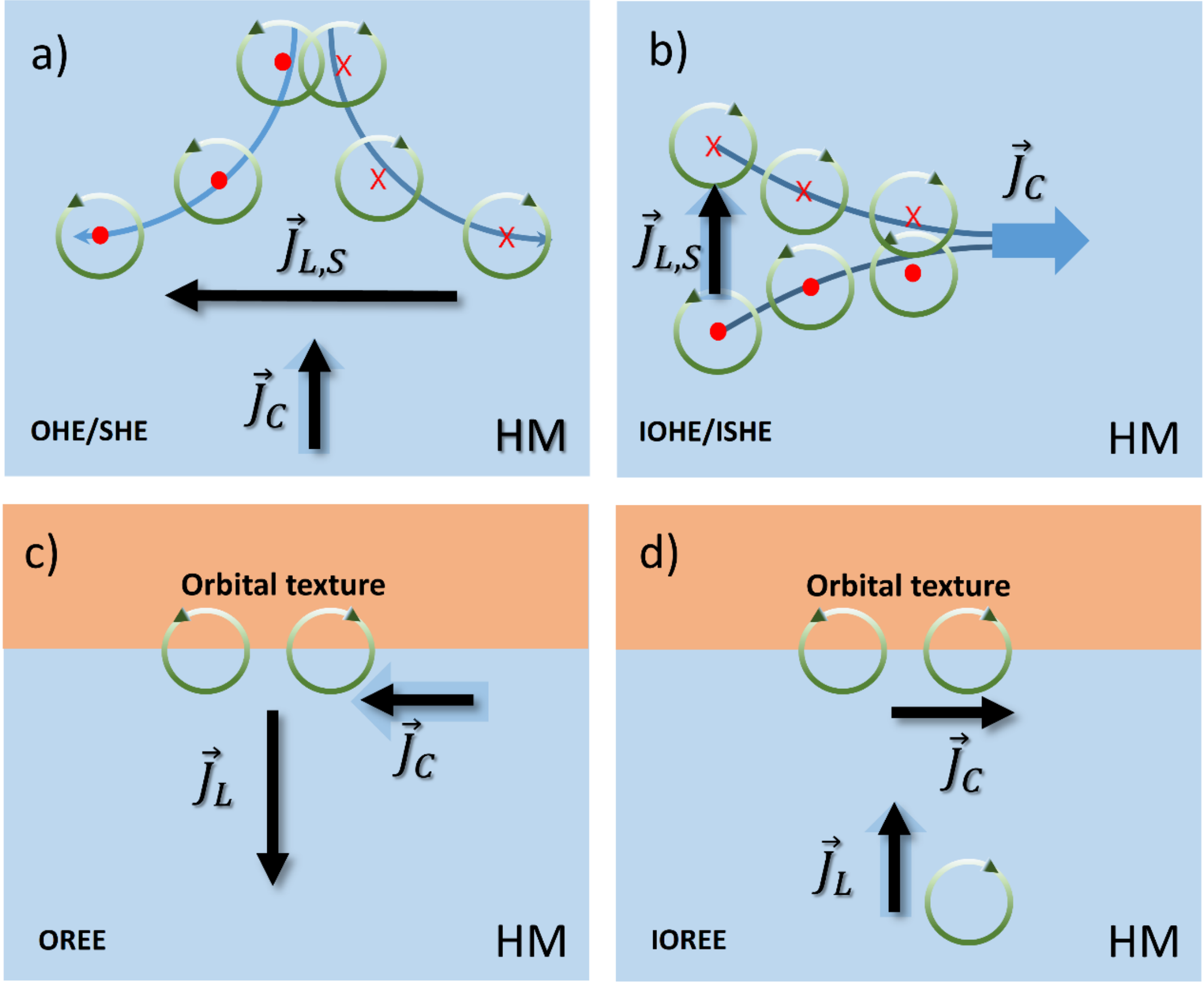

FIG. 1. Schemes illustrating the interaction between charge, spin, and orbital currents in a heterostructure with strong SOC. In the top, (a) and (b), the phenomenon occurs in the volume. In the bottom, (c) and (d), it is driven by the interface. In (a) the direct SHE-OHE is presented, where a charge current $\vec{J}_C$ is converted into a spin-orbital current $\vec{J}_{L,S}$. In (b) the inverse SHE-OHE is presented, where $\vec{J}_{L,S}$ is converted into $\vec{J}_C$. Figs. (c) and (d) illustrate the direct and inverse OREE conversion mechanisms, where the Rashba orbital states are characterized by the orbital textures at the heavy-metal/normal-metal (HM/NM) interface.

Recent works have shown the effectiveness of using light materials to generate enhanced spin-orbital torque transfer in heterostructures. These heterostructures are coated with a thin layer of naturally oxidized $CuO_x$ [16, 17, 20]. This incorporation of light materials into the existing repertoire of spintronic materials has significantly broadened the scope of spin manipulation mechanisms, allowing the use of less expensive materials. The spin-orbital torque enhancement has been demonstrated not only in bulk of light materials, but also in interfaces of Cu/$CuO_x$ driven by OREE. The physics of the OAM phenomena have being clearly demonstrated through several advances, such as the improved damping-like spin-orbit torque (SOT) in Permalloy (Py)/$CuO_x$ [21], enhanced SOT efficiency in thulium iron garnet (TmIG)/Pt/$CuO_x$ [16], and the observation of magnetoresistance driven by OREE in Py/oxidized Cu [17]. In Ref. [16], it is shown that the Pt/$CuO_x$ interface generates an orbital current ($\vec{J}_L$), which then diffuses into the Pt layer. This leads to the emergence of an intertwined spin-orbital current ($\vec{J}_{L,S}$), which subsequently reaches the TIG layer and exerts torque on the local magnetization.

In this study, we performed an extensive investigation on the interplay between spin, orbital, and charge in FM/HM/$CuO_x$, using YIG or Py as FM and Pt or W as HM. Through a comparison of experimental results between FM/HM with FM/HM/$CuO_x$ configurations, we found substantial changes in the ISHE-type signal, suggesting a pivotal role played by HM/$CuO_x$ interface. We observed that W has a different behavior compared to Pt, exhibiting a reduction in the SP signal. We also show that it is important to use HM with strong SOC to generate orbital currents from the injection of a pure spin current. This observation is illustrated in systems like YIG/Ti and YIG/Ti/$CuO_x$, where we observed no differences in the SP signal. This is attributed to the negligible SOC of Ti, meaning that the injection of a pure spin current in Ti does not generate an associated orbital current at the Ti/$CuO_x$ interface, which would produce a charge current. On the other hand, when using Co/Ti samples, we observed a SP signal comparable to that of Co/Pt. This can be explained by the fact that Co injects both spin and orbital current into Ti. Moreover, we show that the IOREE depends exclusively on the polarity of the orbital current, independent of its propagation direction. To investigate this phenomenon, we examined samples of Si/Py/Pt/$CuO_x$ and Si/$CuO_x$/Pt/Py. By extending a phenomenological theory, we adapted it to fit the experimental data by considering orbital angular momentum diffusion in FM/HM bilayers. Certainly, our findings provide insights into the orbital-charge conversion attributed to IOREE, offering potential contributions to the development of electronic devices based on the manipulation of OAM.

## 2. Experimental results

### 2.1 Materials characterization

The YIG films used in this study were grown via liquid phase epitaxy (LPE) on a 0.5 mm thick $Gd_3Ga_5O_{12}$ (GGG) substrate with the out-of-plane axis aligned along the (111) crystalline direction. All other films were deposited by DC sputtering at room temperature with a working pressure of 2.8 mTorr and a base pressure of $2.0 \times 10^{-7}$ Torr or lower. All samples had lateral dimensions of 1.5 x 3.0 mm$^2$, and in all of them the $CuO_x$ layer was obtained by depositing a Copper layer and leaving the samples out in the open air at room temperature for two days. Details about the spin pumping technique used in this investigation can be found in [22].

An investigation of the chemical composition of the GGG/Pt/Cu sample was performed using TEM and atomic resolution energy-dispersive x-ray spectroscopy (EDS). The TEM and EDS results confirmed the existence of an oxidation layer on the surface of the Cu films. Fig. 2 (a) shows the cross-section TEM image of the GGG/Pt/Cu sample interface, where it is possible to distinguish the GGG substrate from the Pt and Cu films. The cap layers of Pt and Au on top of the images were grown afterward during the sample preparation for TEM analysis. To quantify the interfacial chemical diffusion, atomic resolution EDS mapping images were performed on the GGG/Pt/Cu interface areas, and the distribution of each atom element can be seen in Figs. 2(b), 2(c), 2(d), 2(e) and 2(f), corresponding respectively to specific elements: platinum (Pt) is represented by red, gadolinium (Gd) by purple, gallium (Ga) by blue, copper (Cu) by green, oxygen (O) by pink. The atomic percentage of each layer was confirmed by EDS line profile as shown in Fig. 2(g) and 2(h), revealing the presence of the Pt/Cu bilayer spanning a depth range of approximately 10 nm to 63 nm. Note that oxygen is observed in the Cu layer over the range of approximately 53 nm to 63 nm (see Fig. 2(i)). Hence, the TEM and EDS analyses suggest that O atoms diffuse into the Cu layer, implying that the oxidation region in Cu can extend to a depth of up to 10 nm.

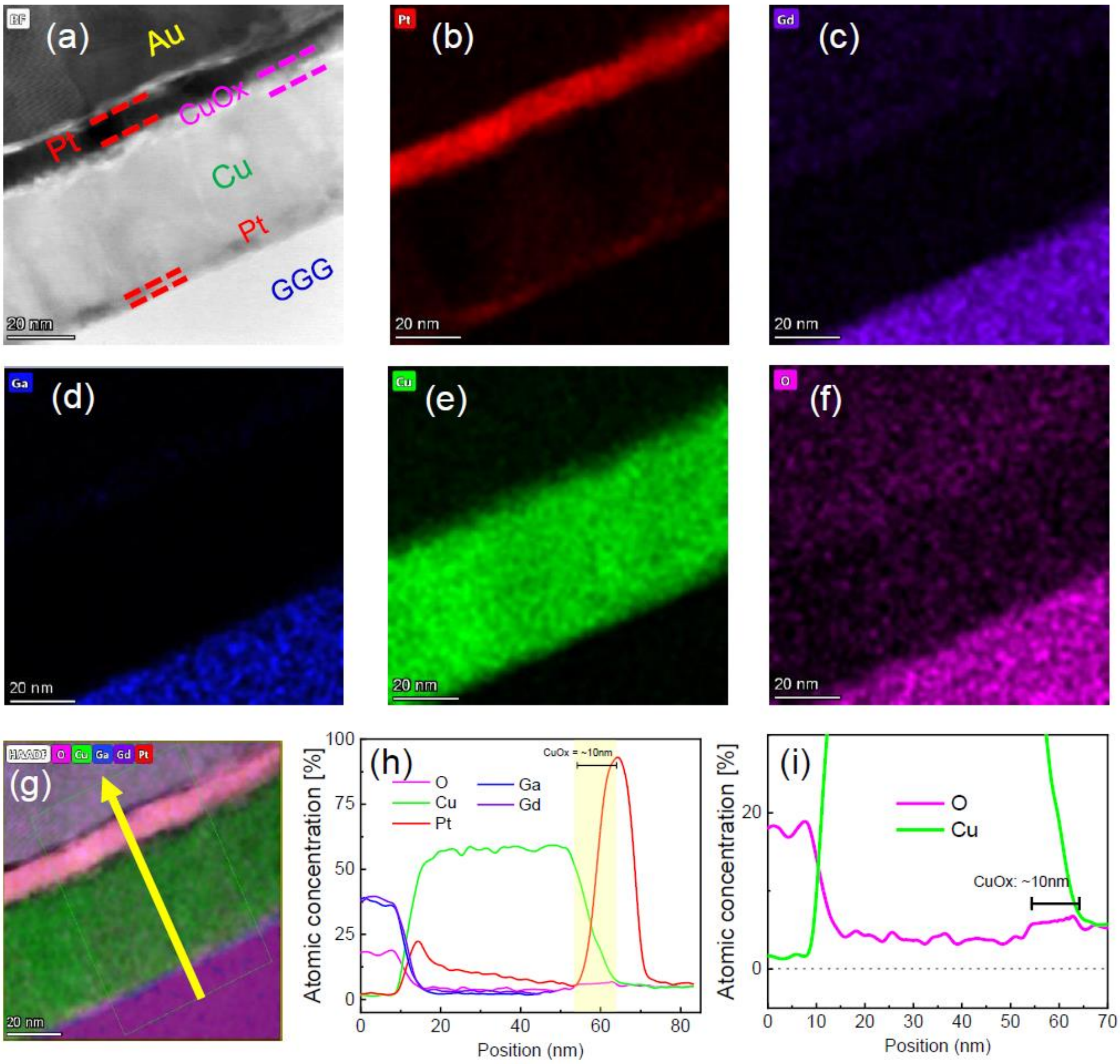

FIG. 2. (a) Cross-sectional TEM image and (b-f) EDS mapping images of the GGG/Pt/Cu sample, displaying chemical element mapping that distinguishes between the GGG substrate (with Ga, Gd, and O elements) and the Pt and Cu films. The color scheme corresponds to specific elements: platinum (Pt) is represented by red, gadolinium (Gd) by purple, gallium (Ga) by blue, copper (Cu) by green, oxygen (O) by pink. (g-i) EDS line scan of atomic fraction of elements Pt, Gd, Ga, Cu and O. The distribution of each atom element is illustrated by their corresponding atomic percentages and the shaded region in (h) indicates the transition area where the O atom diffuses into the Cu layer, displaying a substantial presence of oxygen, with an approximate width of ~10 nm.

### 2.2 Spin pumping in FM/NM/$CuO_x$

The main mechanisms for investigation of the spin-to-charge current interconversion include the SHE and its reciprocal effect ISHE, observed in bulk materials, and the SREE and its reciprocal effect ISREE, observed in systems without spatial inversion symmetry, such as surfaces and interfaces with large SOC. The SHE and ISHE have been extensively investigated in strong SOC materials such as Pt, W, Ta, Pd [11]. Pt is known to have positive spin Hall angle, $\theta_{SH}$, while Ta and W display a negative $\theta_{SH}$ [23-25]. Materials with positive $\theta_{SH}$ exhibit a spin polarization $\hat{\sigma}_S$ parallel to the orbital polarization $\hat{\sigma}_L$, i.e., $(\vec{L} \cdot \vec{S}) > 0$. On the other hand, materials with negative $\theta_{SH}$ present an antiparallel alignment between the spin polarization $\hat{\sigma}_S$ and the orbital polarization $\hat{\sigma}_L$, i.e., $(\vec{L}\cdot\vec{S}) < 0$. In the presence of strong SOC, both orbital and spin effects can occur simultaneously [4], leading to the intertwining of both degrees of freedom. Consequently, the resulting charge current comprises a multitude of effects.

The Spin Pumping (SP) technique [26-28] was employed to investigate the effect on the ISHE and IOHE signals in heterostructures consisting of YIG/W/$CuO_x$ and YIG/Pt/$CuO_x$(3). The charge current resulting from the SP signal, when the intertwined current $\vec{J}_{L,S}$ is considered, is described by the equation $J_C = (2\,e/\hbar)\theta_{eff}J_{L,S}cos(\phi)$, where $\theta_{eff}$ is the effective spin-orbital Hall angle, and the angle between the $\vec{J}_C$ and the voltage measurement direction is given by $\phi$. It is important to note that the effective spin-orbital Hall angle $\theta_{eff}$ must accounts for spin and orbital contributions. In Fig. 3 (a), the typical SP signal for the YIG/Pt(4)/$CuO_x$(3) sample is depicted, where the numbers in parentheses are the layer thicknesses in nm, and the YIG layer thickness is around 400 nm. At $\phi = 0^o$ (blue symbols), a positive sign SP curve is observed, indicating $\theta_{eff} > 0$. At $\phi = 180°$ (red symbols), a sign inversion occurs, while at $\phi = 90°$ the measured voltage is null. The inset of Fig. 3 (a) displays similar results for the YIG/Pt(4) sample, where the peak is much smaller than the signal with a $CuO_x$ cover layer. Fig. 3 (b) presents the results for the YIG/W(4)/$CuO_x$(3) sample, which exhibits an opposite sign compared to Pt, as W possesses $\theta_{eff} < 0$. The same behavior can be observed in the YIG/W(4) sample, as shown in the inset of Fig. 3(b). Additionally, Fig. 3 (c) provides a comparison of the signals obtained with the samples of YIG/W(4)/$CuO_x$(3) and YIG/W(4) at $\phi = 0°$, revealing a reduction in the signal when $CuO_x$ cover layer is added. Fig. 3 (d) shows the behavior of $I_{SP}$ as a function of the thickness of the layer W ($t_W$), which varied from 2 nm to 8 nm. Two sets of samples were prepared: the A series consists of YIG/W($t_w$) (black symbols), while the B series consists of YIG/W($t_w$)/$CuO_x$(3) samples (the red symbols). The B series exhibits a different behavior, where thinner films yield smaller $I_{SP}$ signals, while for larger thicknesses, the $I_{SP}$ tends to approach that of the A series. The solid lines of Fig. 3(d) were obtained by means of the best fit to the experimental data using equations discussed in section 3. The structural characteristics of the W films were analyzed by x-ray diffraction (XRD) measurements and can be seen in Appendix A.

The injection of a pure spin current $\vec{J}_S$ through the YIG/W interface, driven by the precessing YIG magnetization under FMR condition, leads to the intertwining of spin $\vec{S}$ and the angular momentum $\vec{L}$, resulting in the generation of an upward current $\vec{J}_{L,S}$within the W. In this scenario, $\hat{\sigma}_S$ and $\hat{\sigma}_L$ are antiparallel. A portion of this spin-orbital current is subsequently converted into a charge current within the volume of W through the processes of ISHE and IOHE. The remaining orbital current reaches the W/$CuO_x$ interface, where it generates 2D charge current parallel to the interface due to the IOREE. This 2D charge current reduces the original current (as it has the opposite polarity to the bulk charge current), resulting in a smaller signal.

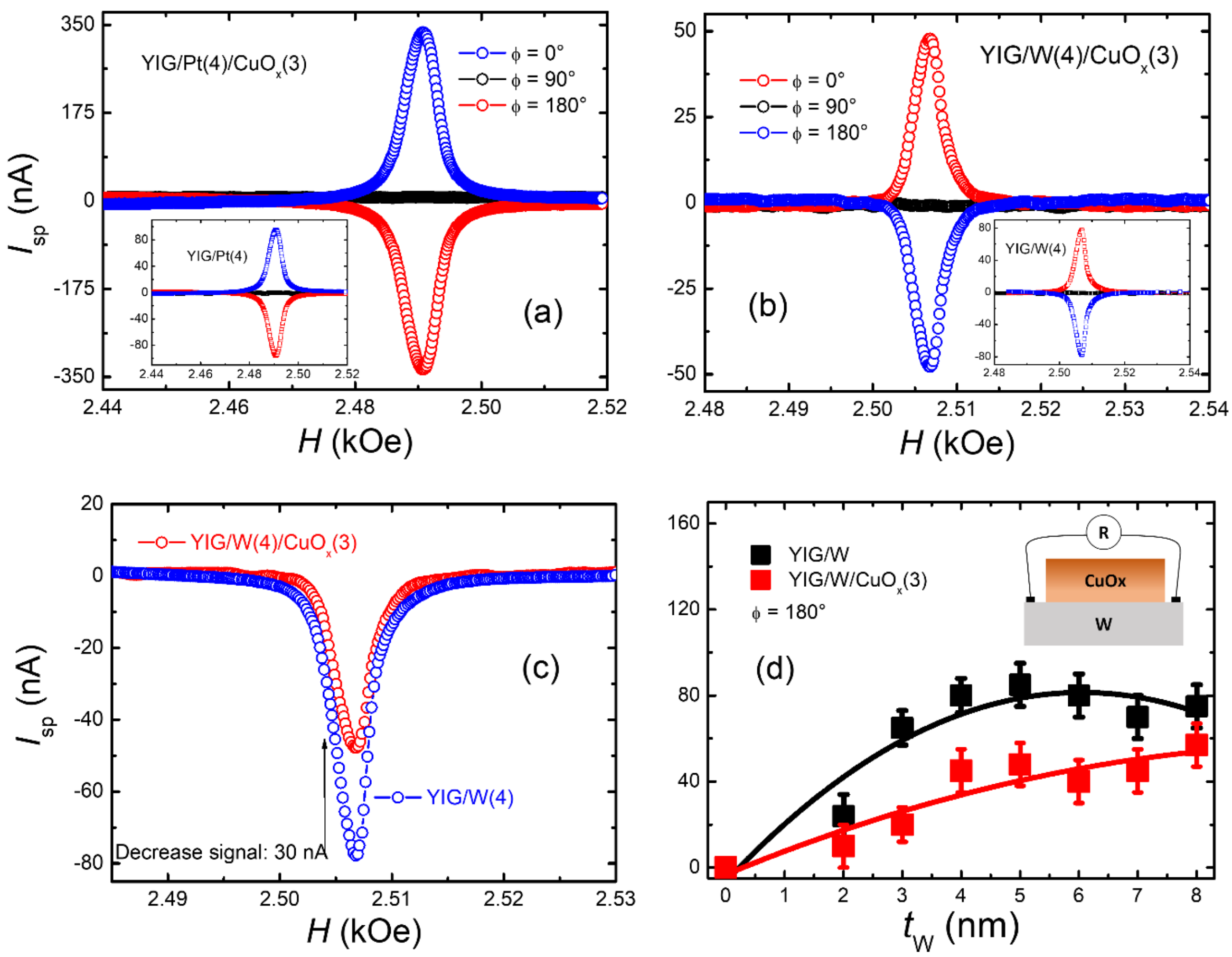

FIG. 3. (a) Presents the typical $I_{SP}$ signals for the samples with and without the CuO$_x$ cover layer (inset) at a fixed $rf$ power of 14 mW and $rf$ frequency of 9.41 GHz. In (a), Pt is used as NM, while in (b), W is used as the NM. These materials exhibit opposite Hall angles, resulting in opposite polarities of the measured signals. (c) Compares the SP signals of the samples with (light blue) and without (dark blue) the CuO$_x$ capping layer. (d) Demonstrates the dependence of $I_{SP}$ on $t_{Pt}$ for the YIG/W($t_W$)/CuOx(3) (red) and YIG/W($t_W$) (black) samples. The solid lines is the theoretical fit.

To gain a better understanding of the role played by SOC in magnetic heterostructures, we fabricated samples of YIG/Ti and YIG/Ti/CuO$_x$. Like the previous experiment, the precessing magnetization generates a spin accumulation at the YIG/Ti interface, which diffuses upwardly as a pure spin current along the Ti layer. The Figs. 4(a) and 4(b) depict the SP signal (for $\phi = 0^{\circ}$) of YIG/Ti and YIG/Ti/CuO$_x$ samples, respectively. Solid lines of Figs. 4(a) and 4(b) depict the respective fits to the experimental data using a Lorentzian function, represented by blue curves ($\phi = 0^{\circ}$) and red curves ($\phi = 180^{\circ}$). Three important pieces of information can be obtained from these data. (i) The weak SP signal generated by Ti has inverse polarity when compared to Pt. (ii) The fit to the experimental data, shown in Figs. 4 (a) and 4 (b), exhibit similar values, meaning that the capping layer of CuO$_x$ practically does not affect the detected signal. (iii) Since Ti exhibits a weak SOC [7], there was almost no generation of orbital current within the material, thus no observable IOREE was detected at the Ti/CuO$_x$ interface. As a result, there was no significant increase in the SP signal when comparing both samples. These findings support the hypothesis that the reduction in the SP signal in the YIG/W/CuO$_x$ samples can be attributed to the orbital effect, particularly the IOREE occurring at the W/$CuO_x$ interface. The key distinction between W/$CuO_x$ and Ti/$CuO_x$ lies in the absence of SOC in Ti, thus the pumped spin current does not convert into a intertwined spin-orbital current inside the material, which leads to no additional signal.

On the other hand, the phenomenon undergoes drastic changes when the FM layer, used to inject the spin current, also injects orbital current (orbital pumping), as observed with Co [29]. Fig. 4(c), shows the symmetric components of the SP data measured in Si/Ti(20)/Co(10) (blue curve) and in Si/Co(10) (red curve). The Co injector, serving as an island, allows for the direct attachment of electrodes on the Ti layer to detect the SP signal. Meanwhile, the measurement in the Co layer captures the self-induced voltage. However unlike YIG, Co has a sizeable SOC [29], thus a spin-orbital current injected into Ti by the Co layer, the pumped spin-orbital undergoes conversion into a charge current by the IOHE, resulting in a strong signal with positive sign. In contrast, the self-induced signal of Co exhibits a weak negative sign. Upon comparing the intensities of the blue signal and the red signal (self-induced voltage at Co), the observed gain is more than eightfold. To compare the SP signal generated by IOHE in Ti with the SP signal generated by ISHE in Pt, we measured SP in Co(12)/Pt(10), as shown in the bottom inset (blue curve). In this case, the SP signal generated by ISHE in Pt is only twice as intense as the SP signal generated by IOHE in Ti. The measurements presented in Figure 4 make it evident that the SP signal generated by orbital pumping in Ti has magnitude comparable with the signal generated by spin pumping in Pt.

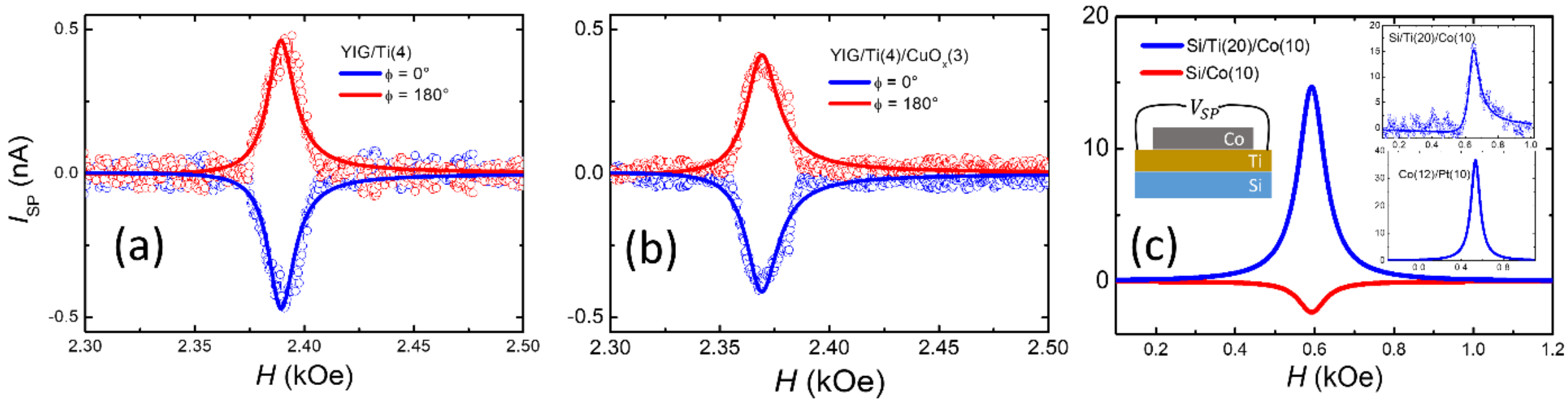

FIG. 4. (a) and (b) shows the SP signals measured in YIG/Ti(4) and YIG/Ti(4)/$CuO_x$(3), respectively. The weak signals were fitted by symmetrical Lorentzian curves, given by the solid lines. Notably, the amplitudes of the signals do not change, indicating that the capping layer of $CuO_x$ does not affect the SP signal. Due to the weak SOC of Ti, no orbital current is being generated within the Ti volume. Solid lines in (c) show the symmetrical component obtained by fitting the data of the SP signal of Si/Ti(20)/Co(10) and Si/Co(10). While the weak SP signal from Si/Co is self-induced, the strong SP signal from Si/Ti/Co is due to the bulk conversion of the orbital current injected into Ti and its conversion by OHE. The bottom inset is the symmetrical part of the SP signal measured in Co(12)/Pt(10).

### 2.3 Spin pumping in all metal heterostructures

Although 3d FM metals such as Fe, Co, Ni, and Py are more versatile and easier to prepare compared to ferrimagnetic insulators, like YIG, these materials exhibit a self-induced SP voltage [30, 31], which can potentially mask the SP signal. This self-induced voltage consists of both symmetric and anti-symmetric components. The anti-symmetric is typically associated with spin rectification effects, while the symmetric component is attributed to spin-Hall like effects [32, 33]. To elucidate the interplay between spin and orbital momenta in metallic heterostructures, we investigate the spin pumping phenomena in two series of heterostructures: series A consists of Si/Py(5)/Pt(4)/$CuO_x$(3) (with and without $CuO_x$ capping layer), while series B consists of Si/$CuO_x$(3)/Pt(4)/Py(5) (with and without $CuO_x$ underlayer). For series B, we initially deposited the copper layer and allowed it to oxidize for two days. Subsequently, we placed the sample back into the sputtering chamber to deposit the Pt and Py films. In series A, the Cu layer, which partially covers the Pt layer, was the final deposition step. Afterward, it was left to oxidize for two days. The only distinction between series A and B is the direction of the spin current injection – upwards for series A and downwards for series B. If the conversion of spin current to charge current is solely given by the inverse SHE, the measured signals should be identical in magnitude but possess opposite polarities. Fig. 5(a) shows the SP signals measured for two samples: Si/Py(5)/Pt(4) and Si/Py(5)/Pt(4)/$CuO_x$(3). In both samples, the spin current is injected upwards through the Py/Pt interface. When comparing the signals obtained from these two samples, a significant increase in the SP signal is observed for the $CuO_x$-coated sample (represented by blue symbols) compared to the uncoated sample (represented by green symbols) at $\phi = 0°$. This enhancement is consistent with previous findings for YIG/Pt/$CuO_x$ [22]. In Fig. 5(a), it is evident that the injected spin current couples with the orbital momentum of Pt, resulting in the generation of intertwined spin-orbital current that propagates upwards until it reaches the Pt/$CuO_x$ interface. At the interface, this spin-orbital current undergoes conversion into a charge current through the IOREE. The converted charge current combines with the bulk charge current, effectively increasing the SP signal. The significant increase is clearly shown in Fig. 5(b), which shows the symmetric component extracted from fitting to the experimental data of Fig. 5(a). When comparing the slopes of the SP signals as a function of the $rf$ power for both samples, as shown in the inset of Fig. 5(b), the sample coated with $CuO_x$ exhibits an increase compared to the uncoated sample. An increase of ~20 nA, for an rf power of 110 mW is shown by the vertical black arrow of Fig. 5(b). However, Fig. 5(d) depicts intriguing results. When the stack order of the layers is inverted, causing the injected spin current from the Py to flow downwards, the SP signal of the sample with an underlayer of $CuO_x$ exhibits a decrease compared to the SP signal of the sample without a $CuO_x$ underlayer. This observation is opposite to the result shown in Fig. 5(a). From the fits to the experimental data, obtained for the symmetric component as shown in Fig. 5(e), the SP signal exhibits a reduction of ~25 nA for the sample with the $CuO_x$ underlayer in comparison with the sample without it. It is important to note in the SP signals of series A and B samples only the spin current injection direction is changed, but the spin polarization is kept identical and, consequently, the orbital polarization remains unchanged.

Our results show that the charge current generated at the Pt/$CuO_x$ interface does not reverse its direction when the spin-orbital current flows from top to bottom. This charge current opposes the charge current generated within the Pt layer, reducing the measured charge current along y direction, as illustrated in Figs. 5(c) and 5(f). The Rashba-type chiral orbital texture present at the Cu/O interface remains unchanged regardless of whether the $CuO_x$ layer is deposited above or below the Pt layer.

Consequently, the charge current generated by the IOREE flows parallel to +y (green arrows at Figs. 5(c) and (f)), while the charge current generated by the spin-orbital current within the Pt layer flows parallel to +y (blue arrow at Fig. (c)) when it is injected from the bottom and parallel to –y (blue arrow at Fig. (f)) when injected from the top. It is important to note that OREE is not affected by the spin current propagation direction and instead depends only on the orbital polarization $\hat{\sigma}_L$. Within the Pt layer, the $\hat{\sigma}_L$ aligns parallel to the $\hat{\sigma}_S$ due to the strong SOC of Pt.

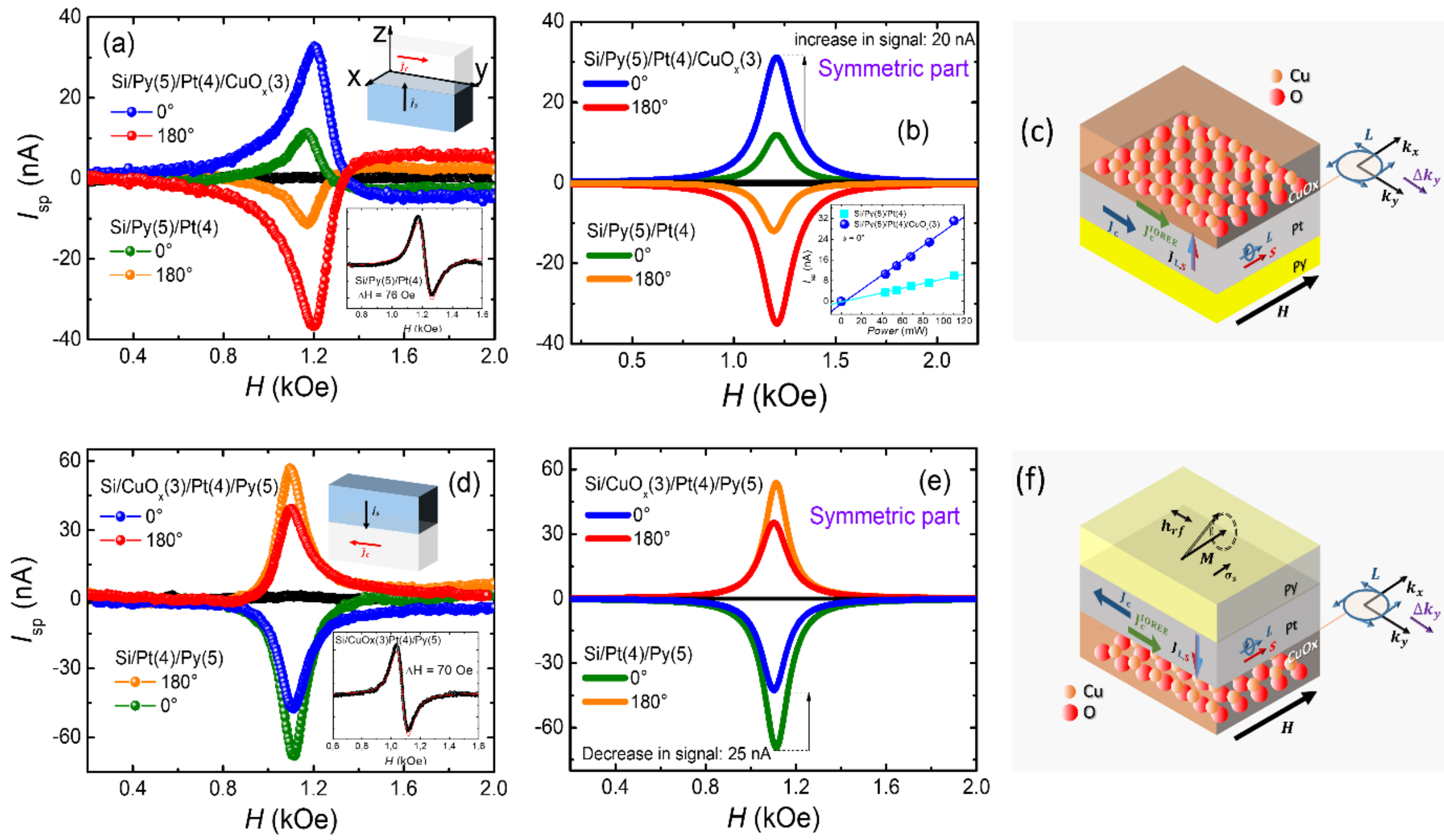

**FIG. 5 -** (a) Typical SP signals for the samples with and without the top layer of $CuO_x$ at $\phi = 0°$. The samples with the top layer of $CuO_x$ are denoted by blue symbols, while those without it are represented by green symbols. The SP signals measured at $\phi = 180°$ have reversed polarities, represented by red symbols (with the top layer of $CuO_x$) and pink symbols (without the top layer of $CuO_x$). The SP data measured at $\phi = 90°$ show no detectable SP signal as expected. Fig. (b) displays the symmetrical component of the SP signal, obtained from fitting the measured data shown in (a) with Lorentzian curves, for samples with and without the $CuO_x$ layer. The inset shows the linear relationship of $I_{SP}$ and $rf$ power. The vertical black arrow represents the increase of the SP signal resulting from the presence of the top layer of $CuO_x$. (d) Typical SP signals for the samples with and without the bottom layer of $CuO_x$ at $\phi = 0°$ (blue and green symbols), and $\phi = 180°$ (orange and red symbols). As the spin current is injected from top, the SP signals exhibit reverse polarity compared to the signals shown in (a). The curves in (e) depict the numerical fittings derived from the data shown in (d) with Lorentzian curves. The vertical black arrow represents the reduction of the SP signal resulting from the presence of the top layer of $CuO_x$. Figs. (c) and (f) illustrate the underlying mechanism responsible for the increase and decrease of the SP signal. In (c), the IOREE and SHE currents are parallel, whereas in (f), they are antiparallel. Insets of Figs. (a) and (d) show the derivative of the FMR absorption signal for the Py layer.

## 3. Phenomenological background

The quantitative interpretation of SHE and OHE presented in this section is based on recently published papers [4,22,34-36]. Basically, the generation of spin and orbital angular momentum currents, along with their interconversion mediated by SOC, can be interpreted in terms of the out-of-equilibrium spin and orbital imbalance, which manifests as a shift in spin and orbital chemical potentials $\mu_S(z)$ and $\mu_L(z)$, respectively. These chemical potentials represent the spin and orbital accumulation, respectively. The accumulation of $S$ or $L$ quantities result in both spin flow and orbital angular momentum flow, and these phenomena can be further analyzed through coupled diffusion equations. A key finding presented in Ref. [4] was the introduction of a coupling parameter, $\lambda_{LS}$, which accounts for the interaction between $L$ and $S$, mediated by the SOC of the material. In Ref. [4], the excitation of orbital current is obtained by applying an electric field, which is different from our approach. Here (and in Ref. [22]), we create a spin accumulation ($\mu_S(z)$), by means of the SP technique, in a material with large SOC, resulting in the simultaneous creation of an orbital accumulation ($\mu_L(z)$). Since materials with large SOC can exhibit two different polarizations of the spin-to-charge conversion processes, such as positive for Pt and negative for W, the time evolution of $\mu_S(z)$ and $\mu_L(z)$ can be expressed as $\mu_L(t) = \upsilon_{LS} C \mu_S(t)$, where $\upsilon_{LS}$ is a variable with only two possible values: $\upsilon_{LS} = \pm 1$, and $C$ is a proportionality constant. In our study, we inject a spin current through the YIG/HM interface, leading to different boundary conditions from Ref. [4] necessary to solve the diffusion equations describing $\mu_S(z)$ and $\mu_L(z)$. In our study, the boundary conditions are given by

$$\begin{cases} \left.\dfrac{d\mu_{S,L}(z)}{dz}\right|_{z=0} = \left(\dfrac{2}{\hbar N D}\right) J_{S,L}(z)\Big|_{z=0} \\ \left.\dfrac{d\mu_{S,L}(z)}{dz}\right|_{z=t_{NM}} = 0. \end{cases} \quad (1)$$

Here, $D$ is the diffusion coefficient, and $N$ represents the density of states per unit volume in the NM layer. To capture the process of spin-to-orbital current conversion, one must add a phenomenological term to spin (orbital) diffusion equation that is proportional to its orbital (spin) counterpart, i.e.,

$$\frac{d^2\mu_S}{dz^2} = \frac{\mu_S}{\lambda_S^2} \pm \frac{\mu_L}{\lambda_{LS}^2} \quad (2)$$

$$\frac{d^2\mu_L}{dz^2} = \frac{\mu_L}{\lambda_L^2} \pm \frac{\mu_S}{\lambda_{LS}^2} \quad (3)$$

where $+$, sign correspond to negative (positive) spin-orbit coupling. To solve the coupled equations (2) and (3) we substitute the former into the latter,

$$\frac{d^4\mu_S}{dz^4} - \left(\frac{1}{\lambda_S^2} + \frac{1}{\lambda_L^2}\right)\frac{d^2\mu_S}{dz^2} + \left(\frac{1}{\lambda_L^2\lambda_S^2} - \frac{1}{\lambda_{LS}^4}\right)\mu_S = 0. \quad (4)$$

The solution of Eq.(4) is

$$\mu_s(z) = Ae^{z/\lambda_1} + Be^{-z/\lambda_1} + Ce^{z/\lambda_2} + De^{-z/\lambda_2} \quad (5)$$

similarly, the equation for $\mu_L$ is obtained, $\mu_L(z) = Ee^{z/\lambda_1} + Fe^{-z/\lambda_1} + Ge^{z/\lambda_2} + He^{-z/\lambda_2}$. The polynomial characteristic leads to,

$$\frac{1}{\lambda_{1,2}^2} = \frac{1}{2}\left[\left(\frac{1}{\lambda_S^2} + \frac{1}{\lambda_L^2}\right) \pm \sqrt{\left(\frac{1}{\lambda_S^2} - \frac{1}{\lambda_L^2}\right)^2 + 4\frac{1}{\lambda_{LS}^4}}\right]. \tag{6}$$

Solving the system of equations, we get the solutions

$$\mu_S(z) = \left(\frac{2}{\hbar ND}\right)\lambda_1 \frac{\left(J_S(0) \mp \frac{J_L(0)}{\gamma_2 \lambda_{LS}^2}\right)}{\left(1 - \frac{\gamma_1}{\gamma_2}\right)} \frac{cosh\,[(t_{NM} - z)/\lambda_1]}{sinh(t_{NM}/\lambda_1)} + \left(\frac{2}{\hbar ND}\right)\lambda_2 \frac{\left(J_S(0) \mp \frac{J_L(0)}{\gamma_1 \lambda_{LS}^2}\right)}{\left(1 - \frac{\gamma_2}{\gamma_1}\right)} \frac{cosh\,[(t_{NM} - z)/\lambda_2]}{sinh(t_{NM}/\lambda_2)} \tag{7}$$

as $\mu_L(z) = Cv_{LS}\mu_S(z)$, then

$$\mu_L(z) = Cv_{LS} \tag{8}$$

where,

$$J_S(0) = \frac{G_S}{e}\mu_S(0), \quad J_L(0) = \frac{G_L}{e}\mu_L(0), \tag{9}$$

$G_{S,L}$is the spin-orbital mixing conductance on the interface FM/HM. The charge current is give by,

$$J_C^{ISHE} = (\hbar/2\,e)\theta_{SH}J_S\sigma_S,$$

and

$$J_C^{IOHE} = (\hbar/2\,e)\theta_{OH}J_L\sigma_L \tag{10}$$

To explain our measured SP signals, the increase in YIG/Pt/$CuO_x$ and a decrease in YIG/W/$CuO_x$, we consider the contributions of both ISHE and IOHE. This allows us to propose a phenomenological equation for the charge current density measured YIG/HM/$CuO_x$ as,

$$J_C = (\hbar/2\,e)\theta_{SH}J_S + (\hbar/2\,e)\theta_{OH}J_L + \lambda_{IOREE}J_L(z = t_{NM}). \tag{11}$$

The first term represents the conversion of the spin component of the intertwined current $J_{L,S}$ into charge current via ISHE within the HM. The second term represents the conversion of the induced orbital current into charge current via IOHE within the HM. This second term can be used, since it arises from the $LS$ coupling, making it analogous to the equation for the ISHE. The third term represents the conversion of the residual orbital current, which reaches the HM/$CuO_x$ interface with Rashba-like states. As a result, the Pt/$CuO_x$ interface exhibits gain in the resulting charge current, while the W/$CuO_x$ interface shows a reduction in the resulting charge current. Therefore, the polarity of the orbital texture of naturally surface-oxidized copper can be modified by changing the HM, leading to an interfacial charge current in the opposite direction to the total charge current. Furthermore, the results presented in Figs. 5 (a) and (d) demonstrate that the IOREE in HM/$CuO_x$(3) remains independent of the direction of the current $\vec{J}_L$. From equations (7-10) it is possible to find the total charge current in the YIG/W and YIG/W/$CuO_x$ samples. The fits to the experimental data using the phenomenological model developed above is presented in Fig. 3 (d).

In conclusion, our investigation of the interaction between spin and orbital currents has yielded significant findings. Through the injection of a pure spin current into a HM layer via the YIG/HM interface, we observed the emergence of orbital momentum accumulation, facilitated by the strong SOC of the HM. This interplay between spin and orbital effects leads to the intriguing phenomenon of transporting orbital angular momentum along the HM layer. As the spin-orbital intertwined $J_{L,S}$ current moves up to the interface of HM/$CuO_x$, there occurs the IREE-like conversion of $J_{L,S}$ into charge current. Moreover, the residual $J_{L,S}$ current that reaches the HM/$CuO_x$ interface is further converted into a charge current by the interfacial IOREE phenomenon. Remarkably, we observed that the charge current generated at the Pt/$CuO_x$ interface exhibits an increase, whereas the charge current at the W/CuOx exhibits a decrease. This aligns with the fact that $(\vec{L}\cdot\vec{S}) > 0$ in Pt, and $(\vec{L}\cdot\vec{S}) < 0$ in W. This result is furthermore confirmed in heterostructure of $CuO_x$/Pt/Py and Py/Pt/$CuO_x$, where the inversion of the layers stack shows a similar behavior. Overall, our work underscores the rich complexity of orbital and spin interactions in HM/$CuO_x$ systems, offering valuable insight into potential applications of spintronics and orbital-based technologies. These compelling findings pave the way for further exploration and innovation in the field of quantum materials and nanoelectronics.

## ACKNOWLEDGMENTS

This research is supported by Conselho Nacional de Desenvolvimento Científico e Tecnológico (CNPq), Coordenação de Aperfeiçoamento de Pessoal de Nível Superior (CAPES), Financiadora de Estudos e Projetos (FINEP), Fundação de Amparo à Ciência e Tecnologia do Estado de Pernambuco (FACEPE), Fundação de Amparo à Pesquisa do Estado de Minas Gerais (FAPEMIG) - Rede de Pesquisa em Materiais 2D and Rede de Nanomagnetismo, INCT of Spintronics and Advanced Magnetic Nanostructures (INCT-SpinNanoMag), CNPq 406836/2022-1 and Chile by Fondo Nacional de Desarrollo Científico y Tecnológico (FONDECYT) Grant No. 1210641 and FONDEQUIP EQM180103. This research used the facilities of the Brazilian Nanotechnology National Laboratory (LNNano), part of the Brazilian Centre for Research in Energy and Materials (CNPEM), a private nonprofit organization under the supervision of the Brazilian Ministry for Science, Technology, and Innovations (MCTI). Therefore, the authors acknowledge LNNano/CNPEM for advanced infrastructure and technical support. The TEM staff is acknowledged for their assistance during the experiments (Proposals No. 20210467 and 20230795, TEM-Titan facility).

## DATA AVAILABILITY STATEMENT

The data that support the findings of this study are available from the corresponding author upon reasonable request.

## APPENDIX A: XRD MEASUREMENTS IN SiOx/W($t_W$)

To obtain structural information of the sputtered W thin films, we performed x-ray diffraction measurements in out-of-plane grazing incident x-ray diffraction (GIXRD). Since in this geometry the substrate signal is almost suppressed, the existence of two distinct crystalline phases (α-W and β-W) as a function of film thickness can be addressed. Fig. 6 shows the GIXRD scans for W films with thickness in the range of 5 nm (purple curve) to 20 nm (orange curve). The vertical blue dashed lines denotes the expected peak positions for (A15) β-W crystalline phase and the vertical red dashed lines denotes the expected peak position for body centered cubic (bcc) α-W phase, according to (JCPDS #03–065-6453)* and (JCPDS #00-004–0806)* crystallographic data, respectively. Also, as can be seen in Fig. 6, for 10 nm W film, the presence of a broad and low intensity peak at 2θ ~ 40° suggests the coexistence of two crystalline phases. Indeed, this peak can be associated to both reflections (210) of β-W phase and (110) of α-W phase, located at 2θ ~ 39.88° and 2θ ~ 40.26°, respectively. On the other hand, for film thickness above 10 nm it is possible to observe three characteristic diffraction peaks. The first (most intense) located at 2θ ~ 40. 44° is closer to the expected position for the reflection (110) of α-W phase and does not exhibit an asymmetrical shape. Furthermore, the other two peaks located at 2θ ~ 58.31º and 2θ ~ 73.42º can only be assigned to (200) and (211) diffraction planes of bcc α-W phase. Taking into account the absence of other β-W phase reflections and that the integrated intensity (area under diffraction curve) of the α -W reflections are increasing with the film thickness, which means that the volume fraction of α -W increases, we can infer that for thickness above 10 nm the films are predominantly α-W phase. Indeed, this fact is in good agreement with previous results that predicts the existence of single α-W phase for thicker films [37]. It is also important to observe in films with a thickness of less than or equal to 10 nm the appearance of peaks between 2θ ~ 52º and 2θ ~ 56º, which are related to the Si/SiO substrate, because the W diffraction peaks have very low intensities.

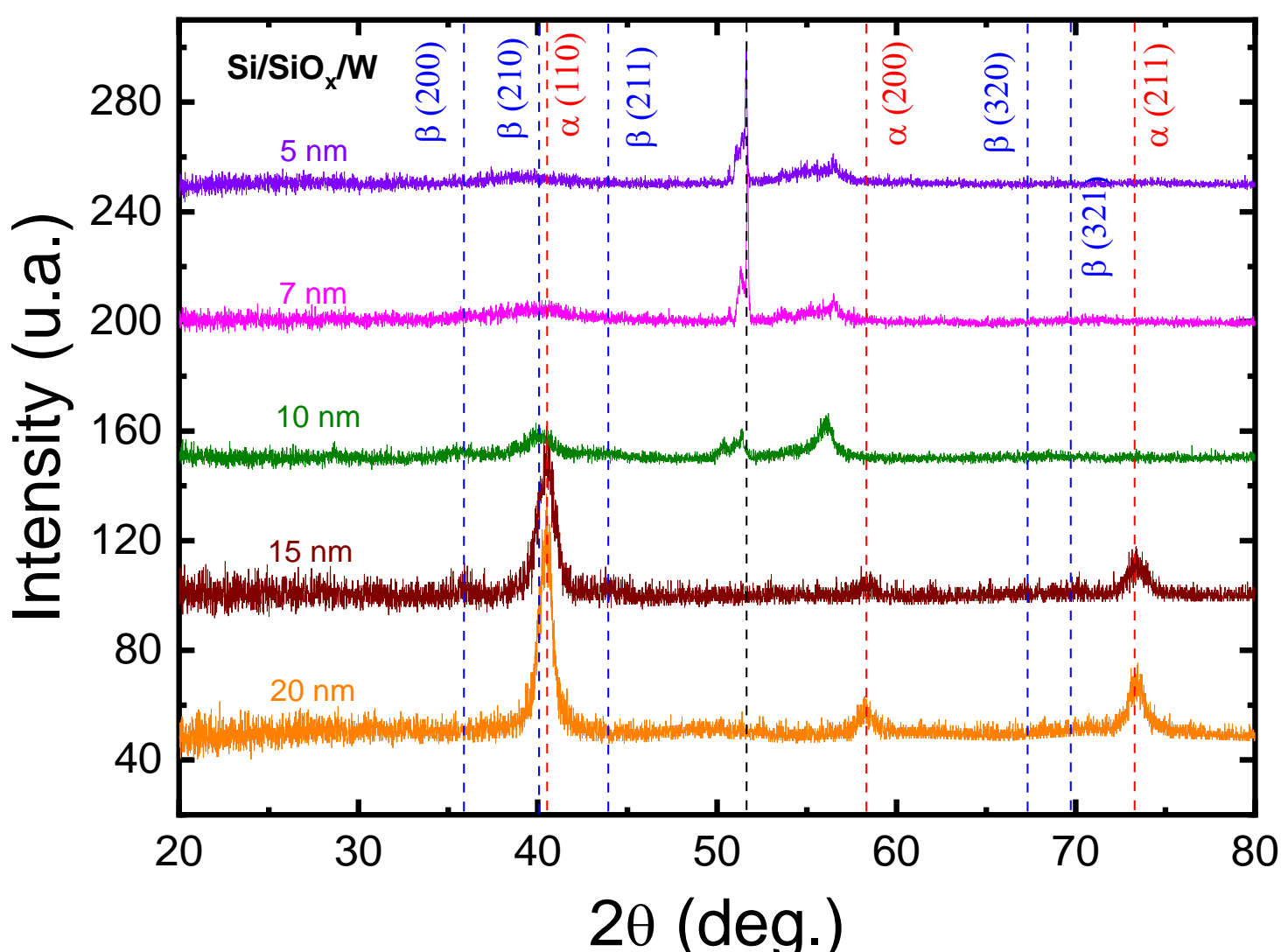

FIG. 6 - Measures of XRD in SiOx/W for different thicknesses of W. For low thicknesses ($t_W < 10nm$) there is a predominance of the $\beta$ phase, while for $t_W > 10nm$ the $\alpha$ phase is predominant.

* Phase identification is made with reference to Powder Diffraction File compiled in International Center for Diffraction Data (ICDD) card system issued by JCPDS (Joint Committee on Powder Diffraction Standards). No. 03–065-6453 for β -W and No. 00-004-0806 for α -W.

## REFERENCES

[1] Dongwook Go, Daegeun Jo, Changyoung Kim, and Hyun-Woo Lee. Intrinsic Spin and Orbital Hall Effects from Orbital Texture. Phys. Rev. Lett. 121, 086602 (2018).

[2] Daegeun Jo, Dongwook Go, and Hyun-Woo Lee. Gigantic intrinsic orbital Hall effects in weakly spin-orbit coupled metals. Phys. Rev. B 98, 214405 (2018).

[3] Soogil Lee, et al. Efficient conversion of orbital Hall current to spin current for spin-orbit torque switching. Commun. Phys. 4, 234 (2021).

[4] Giacomo Sala and Pietro Gambardella. Giant orbital Hall Effect and orbital-to-spin conversion in 3d, 5d, and 4f metallic heterostructures. Phys. Rev. Research 4, 033037 (2022).

[5] Arnab Bose, Fabian Kammerbauer, Dongwook Go, Yuriy Mokrousov, Gerhard Jakob, and Mathias Klaeui. Detection of long-range orbital-Hall torques. ArXiv:2210.02283 (2022).

[6] Hiroki Hayashi, Daegeun Jo, Dongwook Go, Yuriy Mokrousov, Hyun-Woo Lee and Kazuya Ando. Observation of long-range orbital transport and giant orbital torque. ArXiv:2202.13896 (2022).

[7] Young-Gwan Choi, Daegeun Jo, Kyung-Hun Ko, Dongwook Go, Kyung-Han Kim, Hee Gyum Park, Changyoung Kim, Byoung-Chul Min, Gyung-Min Choi, Hyun-Woo Lee. Observation of the orbital Hall effect in a light metal Ti. Nature 619, 52–56 (2023).

[8] Junyeon Kim, Dongwook Go, Hanshen Tsai, Daegeun Jo, Kouta Kondou, Hyun-Woo Lee, and YoshiChika Otani. Nontrivial torque generation by orbital angular momentum injection in ferromagnetic-metal/Cu/Al2O3 trilayers. Phys. Rev. B 103, L020407 (2021).

[9] M. I. Dyakonov, and V.I. Perel. Current-induced spin orientation of electrons in semiconductors. Phys. Lett. A 35 (6), 459 (1971).

[10] J. E. Hirsch. Spin Hall Effect. Phys. Rev. Lett. 83, 1834 (1999).

[11] Jairo Sinova, Sergio O. Valenzuela, J. Wunderlich, C. H. Back, and T. Jungwirth. Spin Hall Effects. Rev. Mod. Phys. 87, 1213 (2015).

[12] Y.-G. Choi, D. Jo, K.-h. Ko, D. Go, and H.-w. Lee. Observation of the orbital Hall effect in a light metal Ti. ArXiv:2109.14847 (2021).

[13] D. Go, et al. Toward surface orbitronics: giant orbital magnetism from the orbital Rashba effect at the surface of sp-metals. Sci. Rep. 7, 46742 (2017).

[14] Annika Johansson, et al. Spin and orbital Edelstein effects in a two-dimensional electron gas: Theory and application to SrTiO3 interfaces. Phys. Rev. Res. 3, 013275 (2021).

[15] Dongwook Go, Daegeun Jo, Tenghua Gao, Kazuya Ando, Stefan Blügel, Hyun-Woo Lee, and Yuriy Mokrousov. Orbital Rashba effect in a surface-oxidized Cu film. Phy. Rev. B, 103, L121113 (2021).

[16] Shilei Ding, Andrew Ross, Dongwook Go, Lorenzo Baldrati, Zengyao Ren, Frank Freimuth, Sven Becker, Fabian Kammerbauer, Jinbo Yang, Gerhard Jakob, Yuriy Mokrousov, and Mathias Kläui. Harnessing Orbital-to-Spin Conversion of Interfacial Orbital Currents for Efficient Spin-Orbit Torques. Phys. Rev. Lett. 125, 177201 (2020).

[17] Shilei Ding, Zhongyu Liang, Dongwook Go, Chao Yun, Mingzhu Xue, Zhou Liu, Sven Becker, Wenyun Yang, Honglin Du, Changsheng Wang, Yingchang Yang, Gerhard Jakob, Mathias Kläui, Yuriy Mokrousov, and Jinbo Yang. Observation of the Orbital Rashba-Edelstein Magnetoresistance, Phys. Rev. Lett. 128, 067201 (2022).

[18] Xi Chen, Yang Liu, Guang Yang, Hui Shi, Chen Hu, Minghua Li, and Haibo Zeng. Giant antidamping orbital torque originating from the orbital Rashba-Edelstein effect in ferromagnetic heterostructures. Nat. Commun. 9, 2569 (2018).

[19] L. Salemi, Marco Berritta, Ashis K. Nandy, and Peter M. Oppeneer. Orbitally dominated Rashba-Edelstein effect in noncentrosymmetric antiferromagnets. Nat. Commun. 10, 5381 (2019).

[20] Zheng-Yu Xiao, Yong-Ji Li, Wei Zhang, Yang-Jia Han; Dong Li, Qian Chen, Zhong-Ming Zeng, Zhi-Yong Quan, Xiao-Hong Xu. Enhancement of torque efficiency and spin Hall angle driven collaboratively by orbital torque and spin–orbit torque. Appl. Phys. Lett. 121, 072404 (2022).

[21] Tenghua Gao, Alireza Qaiumzadeh, Hongyu An, Akira Musha, Yuito Kageyama, Ji Shi, and Kazuya Ando. Intrinsic Spin-Orbit Torque Arising from the Berry Curvature in a Metallic-Magnet/CuOxide Interface. Phys. Rev. Letts. 121, 017202 (2018).

[22] E. Santos, J.E. Abrão, Dongwook Go, L.K. de Assis, Yuriy Mokrousov, J.B.S. Mendes, and A. Azevedo. Inverse Orbital Torque via Spin-Orbital Intertwined States. Phys. Rev. Applied 19, 014069 (2023).

[23] Chunhui Du; Hailong Wang; P. Chris Hammel; Fengyuan Yang. Y3Fe5O12 spin pumping for quantitative understanding of pure spin transport and spin Hall effect in a broad range of materials. J. Appl. Phys. 117, 172603 (2015).

[24] H. L. Wang, C. H. Du, Y. Pu, R. Adur, P. C. Hammel, and F. Y. Yang. Scaling of Spin Hall Angle in 3d, 4d, and 5d Metals from $Y_3Fe_5O_{12}$/Metal Spin Pumping. Phys. Rev. Lett. 112, 197201 (2014).

[25] C. Hahn, G. de Loubens, O. Klein, M. Viret, V. V. Naletov, and J. Ben Youssef. Comparative measurements of inverse spin Hall effects and magnetoresistance in YIG/Pt and YIG/Ta. Phys. Rev. B 87, 174417 (2013).

[26] Yaroslav Tserkovnyak, Arne Brataas, and Gerrit E.W. Bauer. Enhanced Gilbert Damping in Thin Ferromagnetic Films. Phys. Rev. Lett. 88, 117601 (2002).

[27] A. Azevedo, L. H. Vilela Leão, R. L. Rodriguez-Suarez, A. B. Oliveira, S. M. Rezende. Dc effect in ferromagnetic resonance: Evidence of the spin pumping effect? Journal of Applied Phys. 97, 10C715 (2005).

[28] E. Saitoh, M. Ueda, H. Miyajima, G. Tatara. Conversion of spin current into charge current at room temperature: Inverse spin-Hall effect. Applied Phys. Lett. 88, 182509 (2006).

[29] Hiroki Hayashi, Dongwook Go, Yuriy Mokrousov and Kazuya Ando. Observation of orbital pumping. ArXiv: 2304.05266 (2023).

[30] A. Tsukahara, Yuichiro Ando, Yuta Kitamura, Hiroyuki Emoto, Eiji Shikoh, Michael P. Delmo, Teruya Shinjo, and Masashi Shiraishi. Self-induced inverse spin Hall effect in permalloy at room temperature, Phys. Rev. B, 89 (23), 235317 (2014).

[31] A. Azevedo, R. O. Cunha, F. Estrada, O. Alves Santos, J. B. S. Mendes, L. H. Vilela-Leão, R. L. Rodríguez-Suárez, and S. M. Rezende. Electrical detection of ferromagnetic resonance in single layers of permalloy: Evidence of magnonic charge pumping. Phys. Rev. B, 92 (2), 024402 (2015).

[32] A. Azevedo, L. H. Vilela-Leão, R. L. Rodríguez-Suárez, A. F. Lacerda Santos, and S. M. Rezende. Spin pumping and anisotropic magnetoresistance voltages in magnetic bilayers: Theory and experiment. Phys. Rev. B 83, 144402 (2011).

[33] M. Harder, Yongsheng Gui, Can-Ming Hu. Electrical detection of magnetization dynamics via spin rectification effects. Phys. Reports, 661 (23) 1-59 (2016).

[34] Yong Xu, Fan Zhang, Yongshan Liu, Renyou Xu, Yuhao Jiang, Houyi Cheng, Albert Fert, Weisheng Zhao. Inverse Orbital Hall Effect Discovered from Light-Induced Terahertz Emission. Arxiv 2208.01866 (2023).

[35] Tom S. Seifert, Dongwook Go, Hiroki Hayashi, Reza Rouzegar, Frank Freimuth, Kazuya Ando, Yuriy Mokrousov, Tobias Kampfrath. Time-domain observation of ballistic orbital-angular-momentum currents with giant relaxation length in Tungsten. Nat. Nanotechnol. 18, 1132–1138 (2023).

[36] P. Wang, et al., Inverse orbital Hall effect and orbitronic terahertz emission observed in the materials with weak spin-orbit coupling. Npj Quantum Mater. 8, 28 (2023).

[37] Choi, Dooho, et al., Phase, grain structure, stress, and resistivity of sputter-deposited tungsten films. J. Vac. Sci. Technol. A 29, 051512 (2011).